\documentclass[english,aps,prper,reprint,showpacs,titlepage,longbibliography]{revtex4-2}   
\usepackage[utf8]{inputenc}
\usepackage[compatibility=false]{caption} 
\usepackage{booktabs} 
\usepackage{tabularx} 
\usepackage{ragged2e} 
\usepackage{caption} 
\usepackage[T1]{fontenc}	
\usepackage{geometry}
\usepackage{times}
\usepackage{hyperref}  
\usepackage{threeparttable}
\hypersetup{colorlinks=true,urlcolor=blue,citecolor=blue,linkcolor=blue} 
\usepackage{array}
\usepackage{enumerate}
\usepackage{amsmath}
\usepackage{amssymb}
\usepackage{tikz}
\usepackage{graphicx}
\usepackage{multirow}
\usepackage{tcolorbox}
\usepackage{ragged2e}
\usepackage{float}
\usepackage[utf8]{inputenc}

\usepackage[normalem]{ulem}
\usepackage{setspace} 

\begin{document}
\begin{titlepage}

\title{Sense of Belonging and Intent to Persist: Mediating Role of Motivation and Moderating Role of Gender in Physics and Astronomy Graduate Students}

 \author{Swagata Sarkar}
 \affiliation{Department of Curriculum and Instruction, College of Education, Purdue University,
West Lafayette, IN-47907, U.S.A.} 
  
 \author{N. Sanjay Rebello}
 \affiliation{Dept. of Physics and Astronomy / Dept. of Curriculum \& Instruction, Purdue University, West Lafayette, IN-47907, U.S.A.} 


\begin{abstract}
This study investigates how graduate students’ sense of belonging (SB) influences their intent to persist (IP) in physics and astronomy programs, and how this relationship is shaped by the basic psychological needs that drive motivation—autonomy, competence, and relatedness—as well as gender. Grounded in self-determination theory, the analysis treats these three needs as mediators and gender as a moderator. A quantitative survey was administered to graduate students in the Department of Physics and Astronomy at a large public land-grant R1 Midwestern university in the USA. Using probit regressions, we found that SB significantly predicts IP. Autonomy may play a compensatory role when SB is high, competence amplifies the effect of SB on IP, and relatedness buffers against low SB. Gender moderates the relationship: women report lower IP at low levels of SB but exceed men when SB is strong. These findings underscore the importance of fostering a sense of belonging, academic confidence, and social connection—particularly for women in male-dominated STEM fields.

    \clearpage
\end{abstract}

\maketitle
\end{titlepage}
\maketitle
\section{Introduction}\label{sec:introduction}

In higher education, persistence plays a critical role in strengthening the STEM workforce, and fostering persistence is a crucial goal of STEM education. However, considerable barriers persist for STEM graduate students, particularly in physics and astronomy~\cite{porter2019physics}. The graduate program is demanding, requiring students to engage in independent research, coursework, and to contribute to teaching. Persistence enables physics and astronomy graduate students to overcome these challenges, impacting their ability to complete stringent programs and achieve long-term career goals in academia, research, and industry~\cite{porter2024attrition}. Moreover, persistence is associated with career success after graduation. Physicists who demonstrate persistence are more likely to secure positions in academia, industry, or government, as they can navigate the competitive landscapes related to funding and publication~\cite{porter2019physics}.

Behavioral and institutional factors shape the intent to persist in physics and astronomy graduate programs. Affective construct like the sense of belonging~\cite{wang2013motivational} is critical for increasing persistence in the field of physics and astronomy and beyond~\cite{whitcomb2023progression}. However, not much is known about how graduate students experience and cultivate a sense of belonging, and whether that affects persistence in physics and astronomy~\cite{gray2025elements}. These dynamics could be complex and may not always be directly influential. We contend that sense of belonging is highly influenced by a student’s motivation. However, motivation driven by basic psychological needs of which we focus specifically on autonomy, competence, and relatedness as outlined by self-determination theory (SDT)~\cite{ryan2000self}. Therefore, the current study aims to find if motivation in the form of basic psychological needs of autonomy, competence, and relatedness mediates the influence of sense of belonging on the intent to persist in physics and astronomy graduate programs.

Moreover, gender plays a role in the sense of belonging among STEM students, particularly among graduate students of physics and astronomy. Literature shows that women in physics experience a lesser sense of belonging than men, primarily because stereotypes are associated with women’s ability in physics that are responsible for the lower sense of belonging~\cite{stout2013gender}. 

Our research questions are: (i) To what extent does the sense of belonging influence intent to persist among physics and astronomy graduate students? (ii) How do the basic psychological needs that drive motivation mediate, and gender moderate, the relationship between sense of belonging and intent to persist?

\vspace{-5mm}
\section{Literature Review and Theoretical Framework} \label{sec:lit_review}

Persistence in STEM is defined as students' commitment and active pursuit to continue in STEM education, research, and careers, despite the challenges and obstacles that might hinder their smooth progress. Thus, persistence plays a critical role in strengthening the STEM workforce, and fostering this persistence is a crucial goal of STEM education. A blend of personal, social, and institutional factors influence persistence. These impactful factors include a sense of belonging~\cite{wang2013motivational}.  

A sense of belonging is the feeling of being a part of and accepted by a social group~\cite{maslow1943preface}. Sense of belonging is an aspect of students’ mindsets that is getting more recognized by educators and researchers for its influence on motivation, persistence, and significant career decisions of students in STEM~\cite{li2023impact}. Research shows that a sense of belonging is directly related to students’ persistence, and motivation, which drives academic success~\cite{trujillo2014considering, walton2012social}. A strong sense of belonging in STEM also helps mitigate feelings of isolation and promotes sustained interest in challenging fields like physics and astronomy~\cite{hausmann2007sense, binning2020changing}.

Moreover, a strong sense of belonging in STEM can determine women's persistence in STEM careers~\cite{shin2024roles,good2012women}. Thus, raising awareness of the stereotypes and subtle biases, particularly among those who interact with women in STEM, is a potential strategy for improving the persistence of women in these fields~\cite{eddy2016beneath}. A sense of belonging influences overall motivation, which drives persistence in physics and astronomy education and research that contributes to a fast-growing and advanced scientific community and foster innovation and excellence in physics, astronomy, and beyond~\cite{bandura1997self, tinto1997classrooms, strayhorn2012satisfaction, walton2011brief, schunk2009self}.

Motivation is one of the basic psychological needs and a critical driving force for students that predicts self-determination for persistence and future career choice in STEM~\cite{luo2019development, wang2013motivational}. Prior research shows that receiving a degree and achievement in STEM are highly influenced by domain-specific motivational factors, like the sense of belonging~\cite{li2023impact, hansen2024importance}.  A sense of belonging within an academic community helps in fostering intrinsic motivation. Students with an increased sense of belonging engage deeply in their academic environment, enhancing motivation and persistence~\cite{strayhorn2012satisfaction}. According to SDT, an individual’s intrinsic motivation is strongly related to satisfying the three basic psychological needs: autonomy, competency, and relatedness~\cite{ryan2000self}, as it is within the control of students~\cite{deci2012self}. SDT also conveys that the three basic psychological needs, autonomy, competence, and relatedness, steer the motivation continuum in academic and non-academic settings~\cite{ryan2000self}. Intrinsically motivated students are likely to engage in study materials and activities that lead to better learning outcomes and might, likely in turn, increase their sense of belonging within the learning community~\cite{ryan2000self}. Research shows that students who perceive their instructors as supporting autonomy have higher interest and enjoyment in studying physics and experience less anxiety about the subject~\cite{hall2014instructors}. Moreover, students who receive constructive feedback for engaging in challenging tasks succeed because they gain confidence, and in turn enhance their competence in their abilities or performance in physics courses~\cite{kalender2019female}. Physics and astronomy graduate students often encounter social and cultural barriers as they lack relatedness which affects their sense of belonging in an academic environment~\cite{murphy2006girls}. However, positive student–teacher interactions foster relatedness and belonging, which in turn increase persistence in higher education \cite{li2015exploring}. Dost and Mazzoli Smith~\cite{dost2023understanding} also found that physics students are isolated and often encounter stereotypes. In this case, relatedness is mainly associated with a sense of belonging, and it fosters supportive relationships with instructors and peers, which can help identify the barriers underrepresented students encounter in physics. By understanding how the academic environment influences motivation and a sense of belonging, educators can create an inclusive environment to support all students, especially in physics and astronomy~\cite{hazari2007gender}. A strong sense of relatedness can motivate students and enhance their persistence in physics or other STEM fields~\cite{chiu2022school}. SDT offers an effective and practical guide for understanding the dynamics of basic psychological needs and their effect on the sense of belonging in academic settings, such as physics and astronomy graduate programs.

\vspace{-5mm}
\section{Methods}\label{sec:methods}
The goal of this study is to examine how physics and astronomy graduate students experience a sense of belonging within their academic environment and how this, in turn, influences their intent to persist in their graduate programs. Specifically, we investigate the extent to which the basic psychological needs—Autonomy, Competence, and Relatedness—mediate the relationship between Sense of Belonging and Intent to Persist. Additionally, we explore whether gender moderates this relationship, shaping how belonging translates into students’ intent to persist.

\begin{figure}[!htbp]
\centering
\fbox{\includegraphics[width=0.9\linewidth]{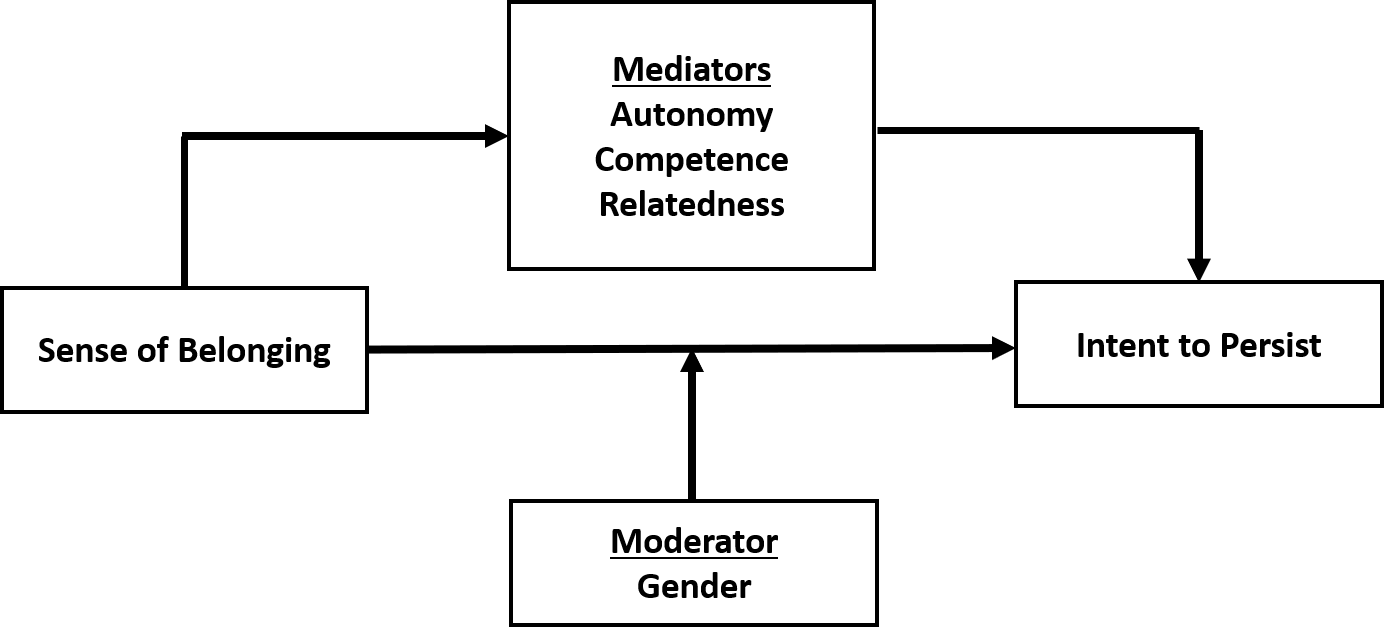}}  
\caption{\justifying{Conceptual model linking sense of belonging to intent to persist with mediators and moderator}}
\label{fig:concept_fmk}
\end{figure}

We conducted a quantitative survey using Qualtrics~\cite{qualtrics2025} with graduate students in the Department of Physics and Astronomy at a large public land-grant R1 university in the Midwest United States. The survey was administered during the Fall 2024 and Spring 2025 semesters, when approximately 178 students were enrolled in the program. Participants received a Qualtrics survey adapted from previously validated instruments \cite{whitcomb2023progression,hausmann2009sense,fedesco2019connections}. The survey included questions on demographics, five items measuring Sense of Belonging \cite{whitcomb2023progression}, two items on Intent to Persist \citep{hausmann2009sense}, and fourteen items assessing basic psychological needs—specifically, four for Autonomy, four for Competence, and six for Relatedness \cite{fedesco2019connections}. All non-demographic items used a six-point Likert scale. A total of 44 students responded to the survey, with 35 completing it in full. In this study, Sense of Belonging is the independent variable, calculated as the average of the five belonging-related items. Intent to Persist is the dependent variable, coded as a binary outcome where 1 indicates intent to persist. The mediators—Autonomy, Competence, and Relatedness—are each measured by averaging responses to their respective items. Gender serves as a moderator and is coded as a binary variable (1 for women, 0 for men). Figure \ref{fig:concept_fmk} presents a conceptual model illustrating the relationship between sense of belonging and intent to persist, as well as the roles of the identified mediators and moderator in shaping that relationship.

\vspace{-5mm}
\section{Findings \&\ Discussion}

To address our first research question—how physics and astronomy graduate students’ Sense of Belonging (SB) influences their Intent to Persist (IP)—we use probit regressions as our primary analytical method. Our use of the probit model rests on two considerations. First, our binary dependent variable, \textit{Intent to Persist}, is better modeled using the latent-variable framework underlying probit, which both constrains predicted probabilities to the 
[0,1] range and captures the inherent nonlinearity of the probability function. Second, among nonlinear binary-choice models, we select probit over logit because our planned extensions involve mediation analysis, where probit is more frequently employed; this choice ensures comparability with prior studies and follows established practices for estimating indirect effects in binary outcome contexts \citep{lee2021point}. Given the small sample size, we opted for this approach instead of more complex econometric models. The probit regression of IP on SB yields a statistically significant slope of 0.489 (see Table~\ref{tab:best_fit_table}), indicating that a stronger sense of belonging is consistently associated with a greater intent to persist. For our second research question, we assess whether this relationship is mediated by three core components of motivation: Autonomy, Competence, and Relatedness; we also examine whether gender moderates the SB–IP relationship. For each motivational factor and for gender, participants are divided into high and low groups based on a median split. Separate probit regression are then estimated for each subgroup to capture the continuous relationship between SB and IP. This method allows us to observe how the strength and direction of the SB–IP relationship vary across different mediators and moderator categories.

\vspace{-5mm}
\subsection{Effect of Motivation Mediators on the SB–IP Relationship}
In this section, we examine how three basic psychological needs that drive motivation—autonomy, competence, and relatedness—influences the relationship between Sense of Belonging (SB) and Intent to Persist (IP). 

\begin{figure}[!htbp]
\centering
\fbox{\includegraphics[width=0.8\linewidth]{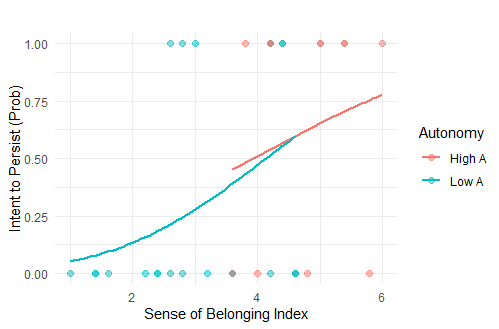}}  
\caption{\justifying{Effect of Autonomy on SB-IP Relationship}}
\label{fig:effect_autonomy}
\end{figure}

\textbf{\textit{Autonomy as a Mediator:}} Figure~\ref{fig:effect_autonomy} shows IP increases with SB across both high and low Autonomy groups. Interestingly, the slope is slightly steeper for students with low Autonomy (0.518) than those with high Autonomy (0.365). Contrary to theoretical expectations, higher Autonomy does not amplify the effect of SB. Instead, these results suggest a compensatory pattern—students with lower Autonomy may rely more on SB to maintain engagement, while those with high Autonomy may be more self-regulated and less influenced by variations in belonging. Autonomy, therefore, does not serve as a traditional enhancing mediator but modulates the SB–IP relationship in a context-dependent way.

\begin{figure}[!htbp]
\centering
\fbox{\includegraphics[width=0.8\linewidth]{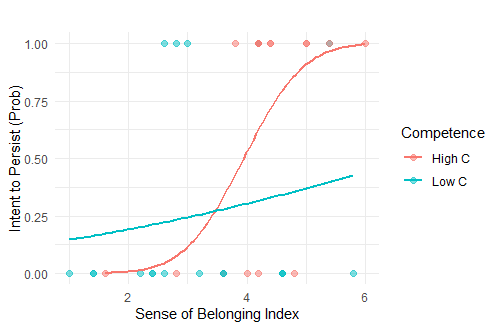}}  
\caption{\justifying{Effect of Competence on SB-IP Relationship}}
\label{fig:effect_competence}
\end{figure}

\textbf{\textit{Competence as a Mediator:}}
Competence emerges as a meaningful and amplifying mediator. Figure~\ref{fig:effect_competence} shows a steep SB–IP slope (1.259) for students with high Competence, indicating that as SB increases, so does their intent to persist. In contrast, students with low Competence show a flatter slope (0.180)—SB has less influence on IP in this group. These findings imply that students who feel both connected (high SB) and capable (high Competence) are much more likely to persist, underscoring the dual importance of social and academic self-efficacy in driving sustained engagement.

\begin{figure}[!htbp]
\centering
\fbox{\includegraphics[width=0.8\linewidth]{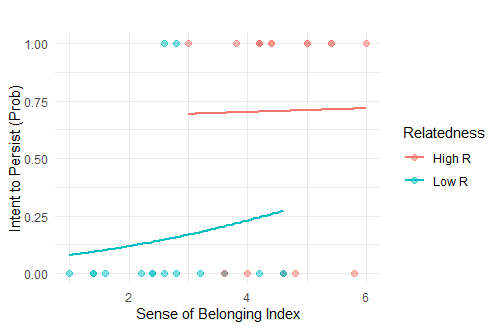}}  
\caption{\justifying{Effect of Relatedness on SB-IP Relationship}}
\label{fig:effect_relatedness}
\end{figure}

\textbf{\textit{Relatedness as a Mediator:}}
Figure~\ref{fig:effect_relatedness} illustrates a buffering mediation pattern for Relatedness. Among students with high levels of Relatedness, IP remains consistently high across the entire range of Sense of Belonging (SB), as indicated by the near-flat slope (0.024). This suggests that strong peer or mentor connections offer a stable foundation for persistence, even when SB varies. In contrast, students with low Relatedness show a much steeper SB–IP slope (0.224), indicating that their intent to persist is more strongly influenced by fluctuations in SB. These findings imply that Relatedness not only mediates the effect of SB on IP but also stabilizes it, shielding students from the adverse effects of low belonging.

\subsection{Effect of Gender as a Moderator of the SB–IP Relationship}

\begin{figure}[!htbp]
\centering
\fbox{\includegraphics[width=0.8\linewidth]{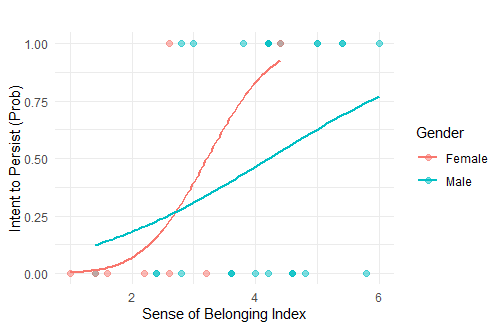}}  
\caption{\justifying{Effect of Gender as a moderator on SP-IP relationship}}
\label{fig:gender_moderator}
\end{figure}

Figure~\ref{fig:gender_moderator} shows that gender moderates the relationship between sense of belonging and IP. The SB–IP slope is substantially steeper for women than for men (1.221 vs. 0.412). At lower levels of SB, women report lower IP than men, suggesting a greater vulnerability to social disconnect. However, as SB increases, women’s IP rises sharply and eventually exceeds that of men, whose IP remains relatively stable throughout the SB spectrum. This pattern indicates that women’s academic engagement is more strongly influenced by their sense of belonging. When SB is low, women are at a disadvantage; when it is high, they surpass their male peers in IP. This asymmetry underscores the pivotal role of SB as a motivating factor for women, particularly in male-dominated fields such as physics and astronomy.

\begin{table}[htbp]
\caption{\justifying {Probit Regression Results for Sense of Belonging and Intent to Persist, with Mediator and Moderator Effects.}}
\begin{tabular}{l|l|l|l} 
\hline \hline
Type                       & Constructs                   & Levels         & Slope           \\ \hline 
                           &                              & Overall Effect & 0.489** (0.199) \\ \hline
\multirow{6}{*}{Mediators} & \multirow{2}{*}{Autonomy}    & High           & 0.365 (0.493)   \\
                           &                              & Low            & 0.518* (0.292)  \\ \cline{2-4}
                           & \multirow{2}{*}{Competence}  & High           & 1.259* (0.705) \\
                           &                              & Low            & 0.180 (0.237)   \\ \cline{2-4}
                           & \multirow{2}{*}{Relatedness} & High           & 0.024 (0.415)   \\
                           &                              & Low            & 0.224 (0.333)   \\ \hline
\multirow{2}{*}{Moderator} & \multirow{2}{*}{Gender}      & Female         & 1.221* (0.823)  \\
                           &                              & Male           & 0.412* (0.248)  \\ 
\hline\hline

\end{tabular}
\begin{tablenotes} \footnotesize\centering
            \item Notes: *p \textless 0.10; **p \textless 0.05; ***p \textless 0.01; \item Standard errors are in parentheses. 
		\end{tablenotes}

\label{tab:best_fit_table}
\end{table}
\vspace{-10mm}
\subsection{Integrated Summary}

Table~\ref{tab:best_fit_table} summarizes the effect sizes from our analyses. Despite a small sample size of 35 respondents, Sense of Belonging (SB) significantly predicts Intent to Persist (IP) at the p < 0.05 level. However, subgroup analyses for mediation and moderation further reduce the sample size, limiting statistical power. Therefore, we emphasize the direction and magnitude of the slopes to identify meaningful patterns in how the SB–IP relationship varies across mediators and moderators.

Among the basic psychological needs that drive motivation, autonomy may play a compensatory role when sense of belonging is high, supporting students’ intent to persist. Competence strengthens the impact of sense of belonging on intent to persist, indicating that students who feel both capable and connected are more likely to continue. Relatedness acts as a buffer, helping students maintain high intent to persist even when sense of belonging is low. Gender also moderates this relationship: women are more sensitive to changes in belonging—showing lower intent to persist when belonging is weak but surpassing men when it is strong.
These findings highlight the need for interventions that strengthen belonging, boost academic confidence, foster social connections, and address gender-specific experiences in male-dominated fields like physics and astronomy.
\vspace{-4mm}

\section{Conclusions \&\ Implications}

This study examines how sense of belonging influences intent to persist in graduate programs, with a focus on the mediating role of motivation and the moderating role of gender, using the framework of self-determination theory. 
In response to our two research questions we found that 1)  a stronger sense of belonging is significantly associated with greater intent to persist in a graduate program;
2) the relationship between sense of belonging and intent to persist is mediated in distinct ways by three basic psychological needs that drive motivation—autonomy, competence, and relatedness—and is moderated by gender. Autonomy plays a compensatory role when sense of belonging is high, competence amplifies the effect of belonging on intent to persist, and relatedness buffers students with low sense of belonging. Gender moderates this dynamic: women’s intent to persist is more sensitive to changes in sense of belonging—lower when sense of belonging is weak, but higher than men’s when it is strong.
The results of our study underscore the need to understand how to support graduate students in physics and astronomy to persist in the graduate program and continue their careers in their respective domains. The literature review highlights that a strong sense of belonging is imperative for increasing persistence and academic achievement in physics and astronomy. By introducing interventions, educators can help create environments that address challenges affecting academic outcomes for physics graduate students.

\vspace{-4mm}
\section{Limitations and Future Work}
This study has certain limitations that also point to avenues for future research. First, the relatively small sample size constrained our ability to use more advanced statistical methods to fully explore the complex relationships between sense of belonging and persistence, as well as the mediating roles of Autonomy, Competence, and Relatedness and the moderating effect of Gender. This constraint was anticipated given the smaller graduate student population in the Department of Physics and Astronomy at the study site. Broader sampling in future work would provide greater analytical depth. Second, results may not be directly transferable to other institutional contexts, as environments and student experiences vary. A multi-institutional sample would help determine whether the observed patterns hold more broadly.
Third, while this study focused on sense of belonging, other affective constructs, such as self-efficacy, science identity, and perceived recognition, may also influence persistence. These could not be included here due to data and statistical power limitations, but future work may incorporate them for a more comprehensive analysis. Finally, future research could pair Self-Determination Theory with critical gender or intersectionality frameworks to examine how structural inequities interact with motivational processes in shaping persistence in physics and astronomy.

Future research could build on these findings by employing a mixed-methods design to examine how affective constructs relate to persistence, with motivation as a mediator.  Following a quantitative survey like ours, scholars may interview those students for the qualitative study. The mixed method design, along with the self-determination theory framework, will help unpack the intersectionality of physics and astronomy graduate students, which might highlight some of their distinct and unique needs.

\section{Acknowledgments}
We thank the survey participants for their responses.

\clearpage
\bibliography{references}
\end{document}